\documentclass[aps, prl, amsmath, floats, floatfix, twocolumn,
superscriptaddress, nofootinbib, showpacs,reprint]{revtex4}
\pdfoutput=1
\usepackage{graphicx}
\usepackage{amsmath,amssymb}
\usepackage{amsfonts}
\usepackage{xspace} 
\usepackage[usenames]{color}
\usepackage{dcolumn}
\usepackage{bm}
\usepackage{mathrsfs}

\usepackage{ulem}
\definecolor {darkgreen}{rgb}{0.2,0.7,0.2}

\newcommand{\eq}{\begin{equation}}
\newcommand{\be}{\begin{equation}}
\newcommand{\eeq}{\end{equation}}
\newcommand{\ee}{\end{equation}}

\newcommand{\mnras}{Mon.\ Not.\ Roy.\ Astron.\ Soc.\ }

\newcommand{\aap}{Astron.\ Astrophys.\ }

\newcommand{\nar}{New Aston.\ Rev.\ }

\begin{document}

\title{Electromagnetic and gravitational outputs from binary neutron star coalescence}

\author
{Carlos Palenzuela$^{1}$, Luis Lehner$^{2}$, Marcelo Ponce$^{3}$, Steven L. Liebling$^{4}$, \\
Matthew Anderson$^{5}$,  David Neilsen$^{6}$,  and Patrick Motl$^{7}$
\\
\normalsize{$^{1}$Canadian Institute for Theoretical Astrophysics, Toronto, Ontario M5S 3H8,
 Canada,}\\
\normalsize{$^{2}$Perimeter Institute for Theoretical Physics,Waterloo, Ontario N2L 2Y5, Canada}\\
\normalsize{$^{3}$Department of Physics, University of Guelph, Guelph, Ontario N1G 2W1, Canada,}\\
\normalsize{$^{4}$Department of Physics, Long Island University, New York 11548, USA}\\
\normalsize{$^{5}$ Pervasive Technology Institute, Indiana University, Bloomington, IN 47405, USA}\\
\normalsize{$^{6}$Department of Physics and Astronomy, 
Brigham Young University, Provo, Utah 84602, USA,}\\
\normalsize{$^{7}$Department of Science, Mathematics and Informatics,
Indiana University Kokomo, Kokomo, IN 46904, USA,}}

\begin{abstract}
The late stage of an inspiraling neutron star binary gives
rise to strong gravitational wave emission due to its highly dynamic, strong
gravity. Moreover, interactions between the stellar magnetospheres can produce
considerable electromagnetic radiation.  
We study this scenario using fully general relativistic, resistive
magneto-hydrodynamics simulations.
We show that these interactions extract kinetic energy from the system,
dissipate heat, and power radiative Poynting flux, as well as develop
current sheets. 
Our results indicate that this power can: (i)~outshine
pulsars in binaries, (ii)~display a distinctive angular- and time-dependent pattern,
and (iii)~radiate within large opening angles.
These properties suggest that some binary neutron star mergers are ideal candidates 
for multimessenger astronomy.
\end{abstract}

\date{}

\maketitle

\noindent{\bf Introduction:} Binary systems involving neutron stars are among
the most likely sources  of detectable gravitational waves~(GW) for detectors such as Advanced LIGO/VIRGO. 
Among other insights
these waves will provide fundamentally new clues about the population
of these systems, constrain the equation of state of matter at nuclear densities, and
provide sensitive tests of general relativity (e.g.~\cite{Sathyaprakash:2009xs}). Additionally, binary neutron
stars~(BNS) are thought to be progenitors of short gamma ray
bursts~(sGRB)~\cite{1984SvAL...10..177B,1989Natur.340..126E,2007PhR...442..166N,2011NewAR..55....1B}
based on energetic, timescale and population considerations.
These sGRBs are extremely energetic, beamed, extra-galactic events that last for less than a couple seconds;
the origin of which has yet to be unambiguously determined.

Models associating BNS with sGRBs involve, at their core, the interaction
of a black hole surrounded by a sufficiently massive disk,
a situation that naturally arises after the collapse of the
hypermassive neutron star resulting from a binary merger.
The interaction of the central compact object with the accretion
disk can power radiation with a hard spectrum and short time
scale characteristic of sGRBs.
Details about how the black hole or disk drives
the radiation, such as via electromagnetic Poynting flux~\cite{2009MNRAS.394.1182K} or 
thermal energy deposition originated by neutrino-antineutrino
annihilation~\cite{2005A&A...436..273A}, remain uncertain.

Correlating observations in both electromagnetic and GW bands
has the potential to
revolutionize our understanding of these systems. 
Examples of what can be gained from such correlations include:
(i)~timing information along with sky localization will test whether 
compact binaries are indeed engines of sGRBS,
(ii)~details from both bands will allow for breaking degeneracies in the
physical parameters (e.g., masses, spins, orbital parameters, etc.) 
of the observed system, and
(iii)~determination of physical parameters 
will clarify the picture of the interaction of the binary with its environment
(e.g.~\cite{Branchesi:2011mi,Kelley:2012tc,lrr-2009-2,Bloom:2009vx}). Additionally, low-latency GW
analysis would allow for localizing a merging binary prior to
the collision itself allowing suitable observatories to be in position to observe
the main event (e.g.~\cite{Aasi:2013wya}).

While obvious candidates for such combined observations are sGRBs, 
it is important to note that not all sGRBs are observable in gamma rays, nor do
all BNS mergers produce sGRBs, if any. 
An exciting possibility, provided by sky localization via gravitational waves,
is the detection ``orphan-GRB'' afterglow signals~\cite{2012ApJ...746...48M,2013MNRAS.430.2121P},
induced by the interaction of the main burst with its environment.
Additionally, EM signals preceding the merger might be detectable.
Such EM precursors would be produced in a relatively cleaner environment,
and so might provide crucial insight on physical parameters
before the complicated, highly non-linear interactions 
expected during the merger epoch. One might also expect the system to radiate at
an opening angle that decreases as the orbit tightens.
Thus, precursors may be identifiable if GW analysis can provide adequate
sky localization.
Tantalizingly, some possible precursors have already been suggested~\cite{2010ApJ...723.1711T}. 

Recently, magnetic interactions 
between the stars~\cite{1996A&A...312..937L,1996Vietri,2001MNRAS.322..695H,Piro:2012rq,Lai:2012qe,Medvedev:2012qf}
and resonant crust cracking~\cite{Tsang:2011ad} have been proposed
as possible precursor mechanisms. Here we focus on the former.
The neutron stars making up the BNS are generally expected to maintain a roughly dipolar magnetic field
and the stars are surrounded by a tenuous, magnetized plasma referred to as a
{\em magnetosphere}.
The interaction of the two stellar magnetospheres coupled with very dynamic gravity can produce
a number of interactions and currents.
The aim of this letter is to study the electromagnetic emission during the pre-merger stage of a BNS, to correlate
it with the emitted gravitational waves, and to 
examine if the interaction of the magnetospheres can yield EM emissions strong enough
to be detected.


\noindent{\bf Physical model:}
We focus on the last orbits of a binary of equal mass,
magnetized, neutron stars in a quasi-circular orbit with initial separation 
$L=45$~km, corresponding to an orbital frequency $\Omega_o=1850$~rad/s.
Each star  has baryonic mass $M=1.62\, M_{\odot}$ and stellar radius $R_*=13.6$~km.
The geometry and matter initial data for this system are obtained with the
LORENE library~\cite{lorene}, assuming a polytropic equation of state
$P/c^2=K \rho^{\Gamma}$ with $\Gamma=2$ and $K=123 G^3 M_\odot^2/c^6$,
which approximates cold nuclear matter. During the evolution
the stars are modeled with a magnetized perfect fluid with an ideal gas
equation of state. Note that the dynamics and interactions of the 
electromagnetic (e.g.~\cite{Ioka:2000yb,2008PhRvL.100s1101A,Lai:2012qe}) and
gravitational (e.g.~\cite{Read:2009yp,Baiotti:2011am}) fields
are largely insensitive to the equation of state during the inspiral.

The stars have an initial, dipolar magnetic field $\bf B$ 
in each star's comoving frame 
described by a magnetic moment $\mu = B_* R_*^3$, with $B_*$ the radial
component at the pole of the star.
To gain insight into the overall behavior of magnetized 
binaries we consider three related initial configurations of the
magnetic moments, with directions specified with respect to the orbital
angular momentum: aligned and equal magnetic moments (case U/U)
with $\mu_1=\mu_2=\mu$, anti-aligned and equal magnetic moments (case U/D)
with $\mu_1=-\mu_2=\mu$, and aligned magnetic moments with 
one-dominant moment (case U/u)  with $\mu_1=100\, \mu_2=\mu$.
In our simulations we set $B_* = 1.5 \times 10^{11}$~G, 
a value on the high end of observations from binaries but still realistic.
The last case~(U/u) has parameters similar to those estimated in 
the double binary pulsar J0737-3039~\cite{2003Natur.426..531B}.

We model this system within a fully consistent implementation that
incorporates general relativity coupled to relativistic, resistive
magnetohydrodynamics in full 3D. Details of our implementation are given
in~\cite{2008PhRvL.100s1101A,Palenzuela:2008sf,Palenzuela:2009yr,
Chawla:2010sw,Palenzuela:2010nf,Lehner:2011aa,Neilsen:2010ax,
2013MNRAS.431.1853P}. 
Our numerical domain extends up to $L=320$~km and contains five
nested fixed mesh refinement~(FMR) grids, each finer grid with twice the
resolution of its parent grid. The highest resolution grid has
$\Delta x = 300$~m and extends up to $58$~km, covering both stars and
the inner part of the magnetosphere. We have also
compared coarser solutions of all the cases
and the results are essentially unchanged.


\noindent{\bf Results:} 
We place particular emphasis on electromagnetic
effects as gravitational phenomena are reasonably well understood for 
this system (for a representative analysis of the late inspiral
GW from this binary see e.g.~\cite{Baiotti:2011am}).
The magnetic field has a negligible effect on 
the orbital dynamics of the system up to 
merger (e.g.~\cite{Ioka:2000yb,2008PhRvL.100s1101A}), as its
contribution to the total inertia is several orders of magnitude 
below that of the matter. The inspiral is
well-described by a post-Newtonian chirp, independent of the magnetic
field.
Consequently, the merger progresses identically for all three cases,
producing the same gravitational signal. The GW luminosity,
to leading order, is
$L_{\rm GW} \simeq 10^{55} \,(M/(2.9 M_{\odot}) )^{10/3} \, (\Omega/\Omega_{\rm ISCO})^{10/3}\,
{\rm ergs/s}$ (with $M$ the total binary mass). We make use of
a fiducial angular frequency $\Omega_{\rm ISCO}=4758$~rad/s, chosen to be that 
of a particle at the inner-most, stable,
circular orbit for a non-spinning black hole of mass $2.9 M_{\odot}$ (this frequency is a good mark of the
onset of the  plunging behavior~\cite{Baiotti:2011am,Buonanno:2007sv}). 
 
Because we focus on the late stage of coalescence, we choose initial data such that
the stars orbit each other for approximately $2.5$ orbits before merging. 
We follow the binary evolution through the merger stage, leaving the post-merger
epoch analysis for future work. (For reference, we set $t=0$ when the stars touch.)

In contrast to the orbital motion of the stars, the behavior of the electromagnetic field
for all three cases
depends sensitively on the orientation of the magnetic moment of the stars.
At a basic level, the accelerated orbital motion of the stars induces only
a small degree of winding on the magnetic fields; thus the magnetospheres
essentially co-rotate with the stars while the magnetic field at the surface
(and the magnetic moment) remains nearly constant until
merger. As each star orbits within the magnetic field of the other,
an electric field and currents are induced.  Such interactions determine the
topology of the resulting field and the net Poynting flux,
as well as other relevant features. In regions where the magnetic field
points in opposite directions, the plasma allows for reconnection of magnetic
field lines, releasing significant energy.
Reconnection can occur within {\em current sheets}, planar regions 
involving almost anti-parallel magnetic field lines supported
by large current densities.

\begin{figure}
\centering
\includegraphics[height=3.5cm,width=9.0cm,angle=0]{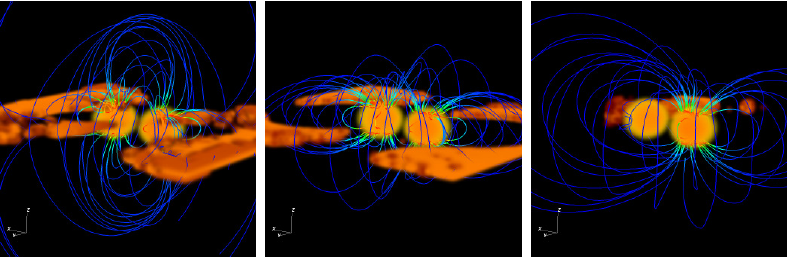}
\caption{Magnetic field configurations (field lines) and
current sheets (orange regions) for  
--from left to right-- the U/D, U/U and U/u cases at time $t=-1.7$ ms.
In all panels, the magnetic field
strength varies from $10^{8}$ (blue) to
$10^{11}$ (red) Gauss. 
The current sheet for the U/U case arises far outside the binary
whereas that for the U/D case arises between the stars
and spirals outward. A trailing dissipation tail is induced
in the U/u case.
\label{fig:fig1}}
\end{figure}

We display (a representative) late-time configuration of the magnetic field lines and
current sheets for the various cases in Fig.~\ref{fig:fig1}.
In the anti-aligned (U/D) and aligned (U/U) cases, a shear layer is induced at
the midplane between the stars, separating two regions filled with magnetically
dominated plasma moving in opposite directions. 
Interestingly, in the U/D case the poloidal component of the magnetic field
points in opposite directions across the midplane, allowing for reconnections.
The resulting magnetic field lines consequently connect 
both stars.
As the stars orbit, these field lines are severely stretched,
increasing their tension and developing a strong toroidal component.
Near the leading edge of each stellar surface,
these field lines undergo a twisting so extreme
that they are bent almost completely backwards, allowing them to reconnect and 
release some of the orbital energy stored by the twisted magnetic fields.

Reconnection between the stars appears absent in the U/U case because the magnetic field points
in the same direction as one moves from one star to the other. 
Instead, far from the shear layer, the configuration resembles the dipole rotator
solution of~\cite{Spitkovsky:2006np} that describes pulsar emission.
This similarity is natural because the system has a net effective dipolar moment at leading order, 
though the symmetry of the binary system
implies an (approximate) periodicity in the solution given
by half the orbital period. 

As the orbit proceeds, both the U/U and the U/D cases
develop current sheets. In the U/D case, they begin between the stars on the orbital plane
and propagate outwards in a spiral pattern. In the U/U case, the current
sheet first arises at far distances. For rotating astrophysical systems, one
defines the {\em light cylinder} as the radius at which the tangential linear
velocity of a co-rotating magnetic field is equal to the speed of light.
It is at the light cylinder for the U/U case that the current sheet first develops 
and continues inward, also with a spiral pattern, as the orbit tightens.

In case  (U/u), 
the magnetic field of the first star eventually dominates that of 
the companion even near its surface. 
Thus, the field is largely described as
an inspiraling dipole perturbed by the induction of the companion.
An interesting effect arises as the magnetic field lines from the strongly 
magnetized star slide off the companion's surface and reconnect behind the star. This reconnection
produces a dissipation tail as illustrated in Fig.~\ref{fig:fig1}. The extent of this tail gradually 
grows as the merger progresses, populating a localized current sheet
behind the weakly magnetized star. 

A qualitative understanding of the radiation from these three configurations,
including the angular distribution, can be obtained from the Poynting flux
shown in Fig.~2.
Both the U/D and U/U cases radiate most strongly along the
shear layer between the two stars and as a consequence their radiation is partially collimated.
Indeed, the flux density in a polar cap  (with opening angle of $\Theta_o<30^o$) is larger
than the average density by factors of $2.5\times$ and $1.9\times$, accounting for $1/3$ and $1/4$ of
the total power, respectively. In contrast, the U/u case radiates mainly on the
equatorial plane and primarily in the direction of the strongly magnetized star
with $2/3$ of the total energy radiated between $60^o<\Theta_o<90^o$.

\begin{figure}
\centering
\includegraphics[height=7.0cm,angle=0]{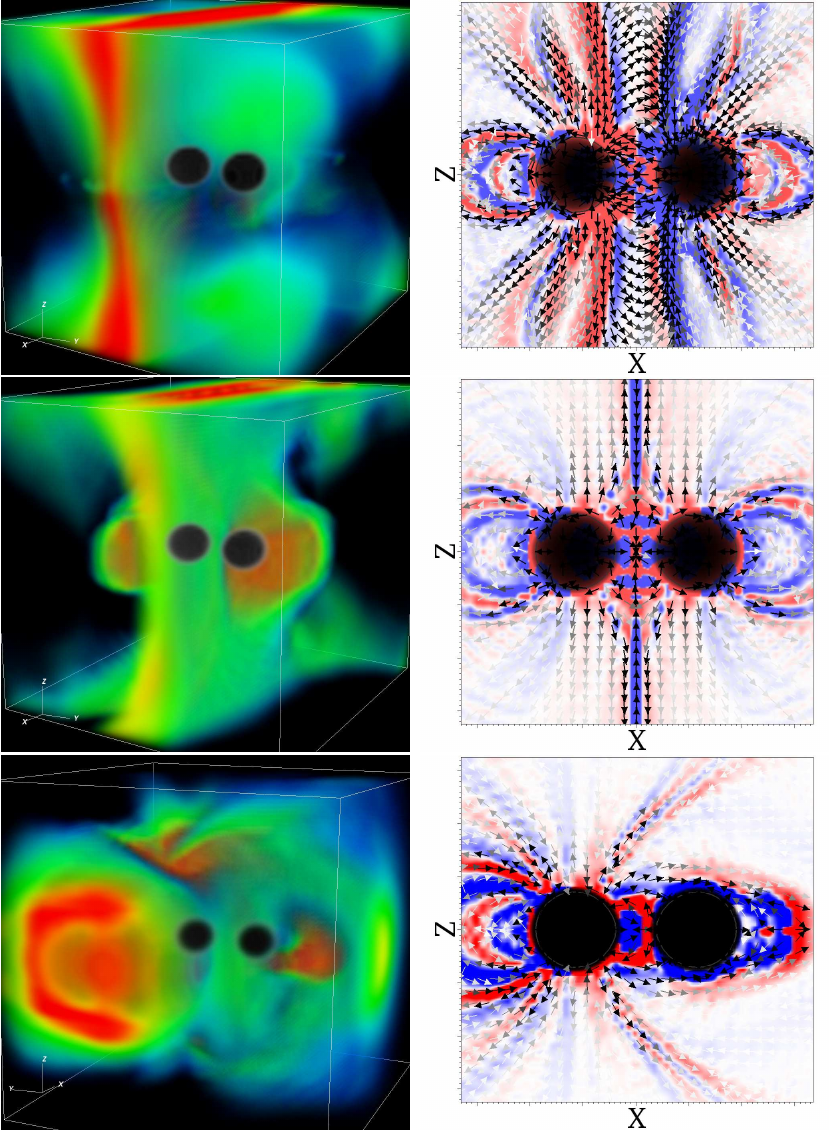}
\caption{Representative snapshots of the electromagnetic energy flux 
at $t=-2.9$ms [left column], and currents~(arrows) and charge density~(negative/positive shaded in red/blue)
 at $t=-0.5$ms [right column];
for the U/D, U/U and U/u cases (top/middle/bottom panels respectively). 
The U/D and U/U cases display currents extending significantly in
both vertical directions, together with EM radiation mainly
directed along the shear layer. In contrast, the currents are mostly
localized in between the stars for the U/u case, with an energy flux
concentrated near the equatorial plane. 
The color scales for the energy flux are arbitrary (with green to red spanning
three orders of magnitude in scale, see Figure 3 for total luminosity vs time for each case). 
For comparison among the three cases we note that the U/D (U/U) case
 is three (two) orders of magnitude larger than the U/u case.
\label{fig:fig2}}
\end{figure}

\begin{figure}
\centering
\includegraphics[height=5.5cm,angle=0]{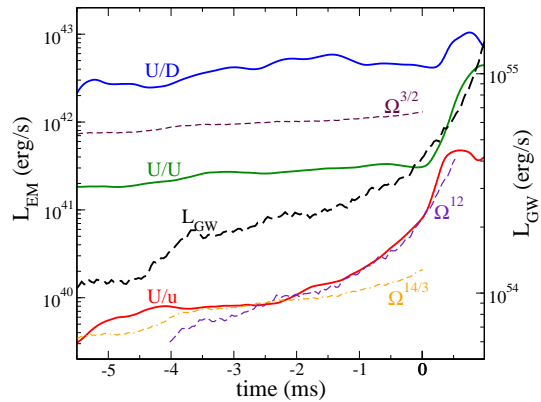}
\caption{Gravitational (right axis) and electromagnetic (left axis) luminosities for the 
three configurations vs time. Three curves illustrating 
$ L \propto \Omega^{p}$ with $p=\{3/2,14/3, 12\}$ are shown as guidance.
\label{fig:fig3}}
\end{figure}

A more quantitative measurement of the electromagnetic radiation of these systems is
provided by integrating the Poynting flux over an encompassing sphere located
at $R_{\rm ext}=180$~km.
Fig.~\ref{fig:fig3} displays this Poynting luminosity as a function of time for the three
configurations.
It is interesting to note that the U/D case is significantly
more radiative than the U/U case. In both cases, the ``inner-engine''
is powered by the magnetic field of each star and by their orbital motion, 
both of which share the same magnitude.
The different luminosities therefore imply a more efficient 
tapping of orbital energy with anti-aligned magnetic moments (U/D)  than
when they are aligned (U/U), possibly due to the 
additional energy radiated
by the release of magnetic tension in the U/D case through reconnections near
the stars.
Although the Poynting flux is generally not directly observable,
this electromagnetic energy can be transferred to kinetic energy of the plasma,
which will radiate through different processes. A detailed
understanding of these processes, even in the context of pulsars, is an active
area of research. We use Poynting flux here as a first approximation to the energetics,
and note that the mechanisms invoked for particle acceleration and emissions from pulsars (e.g. outer-gap, polar-cap, current sheets models)
are applicable here as well.

The Poynting flux pattern and the current sheet structure for all cases
rotate with a periodicity tied to the orbital motion. They thus
trace the spacetime behavior and may help identify the system.


\noindent{\bf Analysis:}
To understand the behavior of the luminosities in
Fig.~\ref{fig:fig3}, in particular their growth as the merger time
is approached, we recall the {\em unipolar inductor} model
of electromagnetic emission~\cite{1969ApJ...156...59G}. 
This model pictures a perfect conductor moving through an ambient magnetic
field, inducing charge separation on its surface and driving currents.
The translational kinetic energy from the moving conductor is extracted in the
form of magnetohydrodynamic waves propagating along the magnetic
field lines~\cite{1965PhRvL..14..171D}.
(We have studied magnetospheric interactions of binary black holes 
and found them well described  by the unipolar behavior~\cite{Palenzuela:2010nf,Palenzuela:2010xn,Neilsen:2010ax}.)
The expected luminosity from this model for a binary system
composed of a magnetized primary star and an unmagnetized
companion (for our masses and radii) is given approximately
by ~\cite{2001MNRAS.322..695H,Piro:2012rq,Lai:2012qe}:
 $L_{\rm ind} \sim 10^{41}\, (B_*/10^{11}G)^2 (\Omega/\Omega_{\rm ISCO})^{14/3}$~ergs/s.

The luminosities can be characterized in terms of powers of the orbital
frequency of the binary as a function of time, $L\propto \Omega^p$ (assuming a constant surface
magnetic field). 
For the U/u case, the early luminosity increases as $\Omega^{14/3}$,
which is consistent with the
unipolar inductor. At later times but still well before the stars touch,
the luminosity increases
much more rapidly, with $p \approx 12$. This slope is significantly
steeper than the usual multipolar emissions. However, this behavior arises
in the most dynamical stage with a rapidly changing multipolar structure, 
and such an analysis need not apply. We defer to future
work a study focused on understanding this behavior.

The U/U and U/D cases differ from the expectation of the unipolar inductor.
At early times,
their luminosity increases with $p\simeq 1-2$ until the
stars come into contact. For these two cases, the transition to
rapid growth of luminosity (again with $p\approx 12$) occurs at later times than the U/u case. 
Interestingly,
the agreement of the slopes for all the cases suggests that the
dynamics near merger is dominated by the formation of the hypermassive neutron
star,
independent of the initial magnetic configuration.


Inspection of the induced currents indicates that
all cases realize an effective circuit (see Fig.~2),  albeit with different
characteristics.
In both the U/U and U/D cases, the circuit extends significantly in both vertical directions, which
contrasts with the more localized currents
in the U/u case.
The shape of the U/u currents roughly resemble those in the U/D case 
running from pole to pole and returning along a mostly
equatorial path
but are much
smaller and restricted to the volume
directly between the stars.
In contrast, the U/U case preserves the symmetry
between the stars and the current leaves the polar regions and returns
along the midplane between the stars.
As a last observation, we note that a significant amount of Joule heating ($J_i E^i$)
is induced and deposited in the plasma between the stars. Relative to the Poynting flux, 
this heating is largest in the U/u case, being  
comparable to its  electromagnetic energy radiated;
for the U/D and U/U cases, on the other hand, the energy dissipated
as heat is roughly 25--50\% of their respective radiated energy. 
We thus stress that these systems display significant differences
with respect to the predictions of the unipolar inductor model.

\noindent{\bf Discussion:} 
We have shown that the interaction of the magnetospheres within a BNS can give
rise to a rich structure that can power strong electromagnetic emissions 
($\simeq 10^{40-43} (B/10^{11}{\rm G})^2$~erg/s) prior to merger.
These luminosities
are at, or higher than, that of the brightest pulsars and would bear
characteristics tied to the orbital behavior.
We have also identified features that can possibly lead to observable
signals tied to the orbital behavior of the system.
The time variability and large opening angle of possible emissions could
help in their detection, especially if already localized in time and space by GW observation. (Binary neutron
stars would spend roughly 30mins in band before merger, allowing for such detection with templates
obtained via PostNewtonian approximations).
Different emission mechanisms are expected near
the current sheets, where strong cooling can give rise to  gamma-rays~\cite{2011SSRv..160...45U,Uzdensky:2012tf} produced
via either synchrotron~\cite{Uzdensky:2012tf} or inverse Compton scattering~\cite{2012arXiv1208.5329L} (see also 
discussion in~\cite{1996A&AS..120C..49A}).  Furthermore, accelerating fields can arise naturally at 
gaps~\cite{1995ApJ...438..314R,2000ApJ...537..964C,Muslimov:2003yz,Muslimov:2004vj}, energizing
a population of particles that emit high energy curvature and synchrotron 
radiation. Understanding which of these mechanisms are the most relevant 
is yet unknown even in pulsar models so there is a large degree of uncertainty in this question.
At a simple level however,
a relativistically expanding electron-positron wind sourced by energy dissipation
and magnetohydrodynamical waves in between the stars could create an X-ray signature~\cite{2001MNRAS.322..695H} preceding or coincident with the merger.
Thus, ISS-Lobster~\cite{Camp:2013cwa} with its high sensitivity and wide field of view
would be very well suited for detecting the associated
electromagnetic counterpart to a binary neutron star merger. 
Depending on how efficiently the Poynting flux is
converted into radiation, these sources could be detectable over a
large fraction of the range of advanced GW detectors.


\noindent{\bf{\em Acknowledgments:}}
It is a pleasure to thank A. Broderick, J. McKinney, A. Spitkovksy for
discussions and Chris Thompson for discussions and comments on the manuscript.
This work was supported by the NSF under grants PHY-0969827~(LIU), PHY-0969811~(BYU),
NSERC through Discovery Grants~(to LL), and NASA award NNX13AH01G.
CP acknowledges support by the Jeffrey L.~Bishop Fellowship.
Research at Perimeter
Institute is supported through Industry Canada and by the Province of Ontario
through the Ministry of Research \& Innovation.  Computations were
performed at XSEDE and Scinet.




\end{document}